\documentstyle[aps,prl,twocolumn]{revtex}
\input epsf

\begin{document}

\title{Periodic thin-film interference filters as one-dimensional photonic crystals}

\author{Dmitry N. Chigrin\thanks{
Author to whom correspondence should be addressed: e-mail: chigrin@uni-wuppertal.de
} and Clivia M. Sotomayor Torres}

\address{Institute of Materials Science, Department of Electrical and Information Engineering\\
University of Wuppertal, D-42097 Wuppertal, Germany}

\maketitle
\begin{abstract}
We review photonic band gap related properties of a simple periodic system of
thin dielectric layers. Properties associated with forbidden and allowed bands
of such one-dimensional photonic crystals are presented. A revision of forbidden
bands properties leads to an \emph{omnidirectional} Bragg mirror design. The
anisotropy of allowed bands suggests the formation of \emph{photon-focusing
caustics} in one-dimensional photonic crystals.
\end{abstract}

\section{Introduction}

Artificial periodically microstructured dielectric materials have attracted
wide\-spread interest in recent years because of the possibility to alter the
dispersion relation of photons \cite{bookJDJ,bookBW95,bookS96}. Such periodic
materials, \textit{photonic crystals}, can exhibit energy gaps of zero density
of states, \textit{full photonic band gaps} (PBG). After the first suggestions
of using photonic gaps to inhibit spontaneous emission \cite{byk72,eli87,john87}
numerous attempts were made to realize three-dimensional (3D) structures that
forbid visible light propagation in all directions (see e.g. \cite{jtl99}).
Much less attention has been focused on the allowed bands of photonic crystals.
The strong dispersion and spatial anisotropy of allowed bands lead to a number
of new optical properties which are inconceivable in conventional crystals \cite{russell86a,zengerle87,kosaka98,kosaka99c}.
In contrast to a full PBG requiring a special topology of the 3D dielectric
lattice, as well as considerably large refractive index contrast of constituents,
strong dispersion and anisotropy of allowed bands are characteristic of any
periodic structures even with a rather small index contrast.

The subject of this paper is the simplest type of photonic crystals, a periodic
arrangement of dielectrics layers of alternating refractive indeces, \( n_{1} \)
and \( n_{2} \), and thicknesses, \( d_{1} \) and \( d_{2} \). Further, we
use subscripts 1 and 2 for low and high index layers, respectively. Such kind
of one-dimensional (1D) photonic crystals belong to the class of the well-known
thin-film interference filters. Thin-film interference filters have proven to
be of great importance in modern optoelectronic applications, ranging from Bragg
mirrors for vertical cavity surface emitting lasers to narrow-band filters for
dense wavelength division multiplexing systems (see e.g. \cite{macleod86,bookFilms87,bookYeh88}).
However, it has been shown recently that some passive devices based on interference
filters can be improved if designed on the basis of photonic crystals \cite{omniMIT2,omniMinsk1,omniBath1,omniMinsk2}.

In general, wave propagation in periodic media can be described in terms of
Bloch waves. For clarity, Bloch waves are briefly discussed in Section II. Two
examples associated with forbidden and allowed bands of a 1D photonic crystal
are presented in Section III and Section IV, respectively. An example of an
interference filter improvement is presented in Section III. Based on photonic
band structure analysis we show how to make a Bragg mirror omnidirectional.
In Section IV, a discussion of some photon-focusing caustics patterns in 1D
photonic crystal is presented. Section V concludes the paper.

\section{Bloch waves\label{secBloch}}

Optical properties of a periodic medium are described by its permittivity, which
is a periodic function of a position in space. In the case of 1D photonic crystals
a medium is periodic in one direction only, so the permittivity
is: \( \varepsilon (z)=\varepsilon (z+l\Lambda  \)), where \emph{z} is the
direction of periodicity, \( \Lambda =d_{1}+d_{2} \) is the period, and \textit{l}
is an integer. According to the Bloch-Floquet theorem (see e.g. \cite{bookYeh88}),
normal electromagnetic modes of such a periodic medium are \begin{equation}
\label{bloch}
\textbf {E}=\textbf {E}_{K}(z)\exp (-iKz),
\end{equation}
where \( \textbf {E}_{K}(z)=\textbf {E}_{K}(z+l\Lambda ) \) is a periodic function
of period \( \Lambda  \). The subscript \textit{K} is the Bloch wave number
and indicates that the function \( \textbf {E}_{K}(z) \) depends on \textit{K}.
The field (\ref{bloch}) can be expanded in Fourier series comprising an infinite
set of partial waves, spatial harmonics. So, a traveling Bloch wave consists
of a group of plane waves propagating together as a stable entity in a particular
direction. To obtain a dispersion relation of an electromagnetic wave in a periodic
medium, relating the frequency of a Bloch wave, \( \omega  \), with the wave
vector, \( \textbf {k} \), a Fourier expansion of (\ref{bloch}) should be
substituted into the Maxwell equations. This results in a dispersion relation
in the form of an infinite system of linear equations. The system should be
properly truncated to get a tractable numerical solution. An exact analytical
solution is generally unavailable. Indeed, a 1D photonic crystal is a unique
system for which an analytical form of the dispersion relation can be derived.
Using the transfer matrix method one can obtain \cite{bookYeh88,russell95}:\begin{equation}
\label{DS_pc}
f(\omega ,\textbf {k})=\cos (K\Lambda )-\left( \frac{1}{2}(A+D)\right) =0.
\end{equation}
 A particular form of the functions \( A(\omega ,\textbf {k}_{\perp }) \) and
\( D(\omega ,\textbf {k}_{\perp }) \) may be found elsewhere~\cite{bookYeh88,russell95},
here \( \textbf {k}_{\perp } \) is the tangential component of the Bloch wave
vector. The dispersion relation given by equation (\ref{DS_pc}) contains all
the information about photonic crystal eigenmodes, describing the properties
of Bloch waves.

In general, the properties of Bloch waves in a periodic structure can differ
drastically from those of a plane waves in an isotropic homogeneous medium.
Well-known dispersion relation of a plane wave in an isotropic homogeneous nondispersive
medium has a form

\begin{equation}
\label{DS_light}
f(\omega ,\textbf {k})=(\omega n/c)^{2}-\textbf {k}^{2}=0,
\end{equation}
where \emph{n} is the refractive index and \emph{c} is the speed of light in
vacuum. A 3D sketch of the dispersion relation (\ref{DS_light}), usually referred
to as the light cone, is depicted in the figure \ref{cones}(a). Only 2D slices
of the wave vector space are presented. The main properties of the band structure
are: \emph{(i) all polarization states are degenerate; (ii) there are not band
gaps;} and \emph{(iii) the direction of the energy flow, the direction of group
velocity, \( \textbf {v}_{g} \), points parallel to the wave vector \( \textbf {k} \).}
The last property follows directly from the dispersion relation (\ref{DS_light}),
because the group velocity is the gradient of the frequency \( \omega  \) in
the wave vector space: \begin{equation}
\label{Vg}
\textbf {v}_{g}=\nabla _{k}\omega (\textbf {k}).
\end{equation}

In a periodic 1D medium, even a small index perturbation yields a band structure
that differs drastically from the light cone. In particular: \emph{(i) The polarization
degeneracy is lifted.} \emph{(ii) Photonic band gaps are developed.} In the
long wavelength limit, the band structure asymptotically follows the light
cone of some homogeneous anisotropic uniaxial crystal, displaying so-called
form birefringence \cite{bookYeh88,bookBorn80}. For higher Bloch wave energies,
new bands, which are separated by frequency and angular gaps appear in the band
structure {[}Fig. \ref{cones}(b){]}. \emph{(iii) The energy flow does not follow
the wave vector direction any more} {[}Fig. \ref{cones}(b){]}. In a periodic
media, the velocity of the energy flow integrated over a unit cell is still
identical to the group velocity \cite{yeh79}. However, due to the strongly
non-circular shape of constant-frequency contours in the wave vector space,
the group velocity (\ref{Vg}) is no longer parallel to the wave vector.

\section{Omnidirectional reflection}

Probably one of the most famous application of thin-film interference filters
is a Bragg mirror, which is a periodic alternation of different dielectric layers
with low and high indices of refraction (Fig.~\ref{Rays}). In general, mirrors
come in two basic varieties: a metallic mirror with a dissipative losses of
few-percent, being in fact an omnidirectional reflector and a dielectric Bragg
mirror. A dielectric Bragg mirror can be made nearly loss-less, but it is highly
reflecting within only limited angular range. A structure, combining the properties
of both mirror types, i.e., being omnidirectional and loss-less, is of strong
interest as it is likely to find many applications in optoelectronics and all-optical
systems.

Until recently, the possibility to design such a {}``perfect mirror{}'' was
mainly associated with 3D photonic crystals having a full PBG. Recently, several
research groups worldwide have reported that a simple-to-fabricate Bragg mirror
suffices to design a low-loss omnidirectional reflector \cite{omniMIT2,omniMinsk1,omniBath1,omniMinsk2}.
This demonstration can lead to various high performance optoelectronic devices
employed at any desirable wavelength. Efficient antenna substrates, energy saving
filters, enclosures for microcavities \cite{omniCavity1} and waveguides \cite{omniGuide1}
are a few potential applications.

The properties of Bloch waves inside a Bragg mirror are governed by the dispersion
relation (\ref{DS_pc}). Due to the planar geometry of the problem, the separation
of the electromagnetic field into TE (transverse electric) and TM (transverse
magnetic) polarization states is possible, where the electric or magnetic field
vector is parallel to the layers interfaces, respectively. This splits the problem
of light interaction with a Bragg mirror into two independent problems.

A projected band structure of an infinite periodic system of layers is depicted
in the figure~\ref{PBGk}. This is a projection of a 3D band structure {[}Fig.
\ref{cones}(b){]} onto the \( \omega -k_{x} \) plane, where \( k_{x} \) is
the tangential component of the wave vector \( \textbf {k} \), assuming that
the plane of the incidence is \( x-z \). The refractive indices, \( n_{1}=1.4 \)
and \( n_{2}=3.4 \), are chosen close to ones of SiO\( _{2} \) and Si in the
near IR region. Thicknesses of the layers are equal (\( d_{1}=d_{2} \)). The
top panel is for TE polarization, and the bottom one for TM. An infinite periodic
structure can support both propagating and evanescent Bloch waves. In figure~\ref{PBGk},
gray areas correspond to the propagating states, whereas white areas contain
the evanescent states only and are referred to as photonic band gaps.

When the frequency and the wave vector of a wave, impinging externally at an
angle, \( \alpha _{inc} \), from a homogeneous medium of refractive index,
\( n \), onto a thin-film stack, lies within the band gaps, an incident wave
undergoes strong reflection. Pronounced high reflection bands (stopbands), depend
strongly on frequency and angle of incidence. These can be easily understood
from figure \ref{PBGk}. Photonic band gaps \textit{(i) rapidly move to higher
frequencies with increasing incident angle, denoted by the increase of the tangential
component of the wave vector}; \emph{(ii) the TM band gaps tends to zero when
approaching the Brewster light-line}, where \( \omega =c\left| {\textbf {k}}_{\perp }\right| /n_{1}\sin {\alpha _{B}} \)
(Fig.~\ref{PBGk}), \( \alpha _{B}=\arctan n_{2}/n_{1} \) is the Brewster
angle. The TM polarized wave propagates without any reflection from \( n_{1} \)
to \( n_{2} \) layer, and from \( n_{2} \) to \( n_{1} \) layer, at the Brewster
angle \( \alpha _{B} \). These properties of the band structure restrict the
angular aperture of a polarization insensitive range of high reflectance.

In essence, omni-directional reflectance can be achieved due to the limitation
of the number of modes that can be excited by externally incident waves inside
the Bragg mirror. Light coming from the low-index ambient medium \( (n<n_{1},n_{2}) \)
is funneled into the internal cone narrowed by Snell's law (Fig. \ref{Rays}).
Angles inside the crystal should be so small as to have band gaps open up to
the grazing incident angles. In particular, \emph{(i) sufficiently large index
contrast of the layers with respect to the ambient medium ensures that light
coming from the outside will never go below the Brewster's angle inside the
crystal} (Fig. \ref{Rays}) \emph{}and \emph{(ii) sufficiently large refractive
index contrast of the layers themselves can keep the band gaps open up to the
grazing angles} \cite{omniMIT2,omniMinsk1,omniBath1,omniMinsk2}\emph{.}

A reduced region of k-space, where electromagnetic modes of the photonic crystals
can be excited by externally incident wave, lies above the light-line (Fig.~\ref{PBGk}),
which is a 2D projection of the light cone of an ambient medium. For example,
in the case of the Si/SiO\( _{2} \) structure in air the first two band gaps
are open for all external angles of incidence (shaded areas in figure~\ref{PBGk}).
No propagating mode are allowed in the stack for any propagating mode in the
ambient medium for either polarizations. \emph{}Thus, total omnidirectional
reflection arises.

To demonstrate an omnidirectional Bragg mirror, transmission spectra of light
impinging at normal and oblique (30\( ^{\circ } \), 60\( ^{\circ } \), 89\( ^{\circ } \))
angles of incidence onto the Si/SiO\( _{2} \) structure are shown in figure
\ref{trans}. Here we assume for calculations that \( d_{2}/(d_{1}+d_{2})\approx 0.324 \)
\cite{omniMinsk2}. At normalized frequencies 0.25--0.32, a high reflectivity
can be observed at all angles of incidence for both fundamental polarizations.
Such a Bragg mirror can be treated as an omnidirectional high reflector with
a relative bandwidth of about 25\%. To obtain, e.g., an omnidirectional reflection
centered at the radiation wavelength \( \lambda =1.5\mu m \), one needs a structure
with a period of about \( 0.412\mu m \).

\section{Photon-focusing caustics}

The strongly non-spherical shape of constant-frequency surfaces of 1D photonic
crystals {[}Fig.~\ref{cones}(b){]} leads to an anisotropic energy flux. We
will refer to this phenomenon as to \emph{photon-focusing}, in analogy to \emph{phonon}-\emph{focusing},
which takes place for ballistic propagation of phonons in solids \cite{bookWolfe98}.

The anisotropy of periodic multilayer structures in the long-wavelength regime
(\( \lambda \gg \Lambda  \)) is a well known phenomenon \cite{bookBorn80},
form birefringence. In the limit of long wavelengths, constant-frequency surfaces
of electromagnetic modes in a photonic crystal are similar to the constant-frequency
surfaces of a negative uniaxial crystal. At shorter wavelengths, the anisotropy
becomes stronger and constant-frequency surfaces become distorted, especially
near the boundaries of the Brillouin zone {[}Fig. \ref{cones}(b){]}.

The energy flow inside a photonic crystal is generally not along the direction
of the wave vector. The energy propagates along the direction of the group velocity
(\ref{Vg}), i.e., along the outward normal to the constant-frequency surface.
In fact, even a small perturbation of the refractive index suffices to distort
drastically a constant-frequency surface from its original spherical shape.
As a consequence, an isotropic distribution of wave vectors emanating from an
isotropic photon source inside the crystal does not imply an isotropic distribution
of energy flux.

To plot a constant-frequency surface of 1D photonic crystals, the dispersion
relation (\ref{DS_pc}) should be solved at a fixed frequency \( \omega  \).
In particular, one should vary the value of the tangential component of the
wave vector \( \textbf {k}_{\perp } \) and solve equation (\ref{DS_pc}) for
a Bloch wave number \( K \), which is basically the \( k_{z} \) component
of the Bloch wave vector. In figure \ref{ds_Caustics} such a constant-frequency
contour, which is an intersection of a constant-frequency surface by the \( K-k_{x} \)
plane, is depicted. To produce a 3D plot of a constant-frequency surface it
is necessary to solve equation (\ref{DS_pc}) for all wave vectors \( \textbf {k}_{\perp } \)
in the \( k_{x}-k_{y} \) plane. However, 1D photonic crystals are invariant
under any rotation around the normal to the layers boundaries. That is why,
a constant-frequency surface of a 1D photonic crystal is simply a revolution
surface of a 2D constant-frequency contour calculated in any plane perpendicular
to the boundaries of the layers.

We plot a constant-frequency contours of a homogeneous isotropic (dashed curve)
and a slightly modulated periodic multilayer (solid curve) media in figure \ref{ds_Caustics}.
The refractive index of the homogeneous medium is \( n=3.4 \) and a perturbation
leads to a structure with indeces \( n_{1}=3.4 \) and \( n_{2}=3.41 \). We
further assume that layers thicknesses are equal (\( d_{1}=d_{2} \)). Constant-frequency
contours are presented for the normalized frequency \( \omega =0.146 \), which
is within a first band gap of the periodic medium. Near the Brillouin zone boundary
the constant-frequency contour of photonic crystal is strongly non-circular.
It consists of a concave and a convex regions with positive and negative Gaussian
curvatures, respectively. The curvature of a photonic crystal constant-frequency
contour is depicted in the inset to figure \ref{ds_Caustics}. In contrast to
the constant curvature of a homogeneous medium (dashed line in the inset) the
curvature of a photonic crystal displays a strong variation. There is an inflection
point where the curvature vanishes (point A in Fig. \ref{ds_Caustics}).

It is known from phonon imaging experiments \cite{bookWolfe98}, that, in the
geometrical optics approximation, a vanishing curvature of the constant-frequency
surface leads to an infinite energy flux from a point source along the corresponding
group velocity direction. The geometrical optics approximation assumes that
phonon (photon) wavelength is much smaller than the source and detector sizes,
and certainly much smaller than the distance between the source and the detector.
These sharp singularities in the energy flux are called \emph{phonon-focusing
caustics} \cite{bookWolfe98}.

Because the energy flux is inversely proportional to the curvature of the constant-frequency
surfaces \cite{bookWolfe98}, the points with zero curvature (Fig. \ref{ds_Caustics})
will lead to \emph{photon-focusing caustics} of electromagnetic energy flux
inside the 1D photonic crystal. Due to the rotation invariance of a photonic
crystal, the Gaussian curvature of the constant-frequency surface vanishes along
the circle produced by the zero curvature point A (Fig. \ref{ds_Caustics}).
If an isotropic non-coherent light source is places inside a thick slab of 1D
periodic medium, then sharply defined circular peaks in the intensity distribution
should be detected outside the sample.

To substantiate this prediction we present a Monte Carlo simulation of an angular
intensity distribution which is due to an isotropic point source inside the
1D photonic crystal. Our Monte Carlo intensity diagram construction process
consists in generating a uniform random distribution of wave vectors \( \textbf {k} \)
and solving equations (\ref{DS_pc}) and (\ref{Vg}) to obtain group velocities
of the electromagnetic modes belonging to each value of \( \textbf {k} \).
Then the group velocity vectors are collected in all directions to form an angular
plot of intensity.

The angular distribution depicted in figure \ref{int_Caustics} was generated
with a large number (\( 10^{6} \)) of initial wave vectors \( \textbf {k} \)
and a {}``source{}'' situated inside the photonic crystal. The intensity was
collected by a {}``detector{}'' placed on the surface of the crystal along
one of the layers boundary. An intense peak appears near the direction corresponding
to the inflection point with zero curvature (point A in Fig. \ref{ds_Caustics}).
This direction corresponds to \( 3.6^{\circ } \) out of the normal to the crystal
boundary. Figure \ref{int_Caustics} shows how the light intensity diminishes
rapidly with increasing angular deviation from that direction. At observation
angles around zero the vanishing intensity is due to the band gap of the photonic
crystal.

\section{Conclusion}

In conclusion, we have presented a brief review of the properties of periodic
thin-film interference filters, 1D photonic crystals. We have discussed an improvement
of a Bragg mirror design. We have explained how to make a Bragg mirror omnidirectional.
Finally we have predicted the formation of photon-focusing caustics in 1D photonic
crystals. Being the simplest type of photonic crystals, thin-film interference
filters, should be a good laboratory structure to study photonic band gap related
phenomena as well as a good candidates for improvement and design of optoelectronic
devices. For example, problems of antenna on top of an omnidirectional mirror
and microcavity with omnidirectional Bragg mirrors are interesting for further
study.

\section*{Acknowledgments}

DNC wishes to dedicate this paper to his teacher and friend Andrei V. Lavrinenko,
who introduced him to the world of electromagnetism of complex materials. This
work was partially supported by the EU-IST project PHOBOS IST-1999-19009.


\center

\noindent \epsfxsize=8.5cm \epsfbox{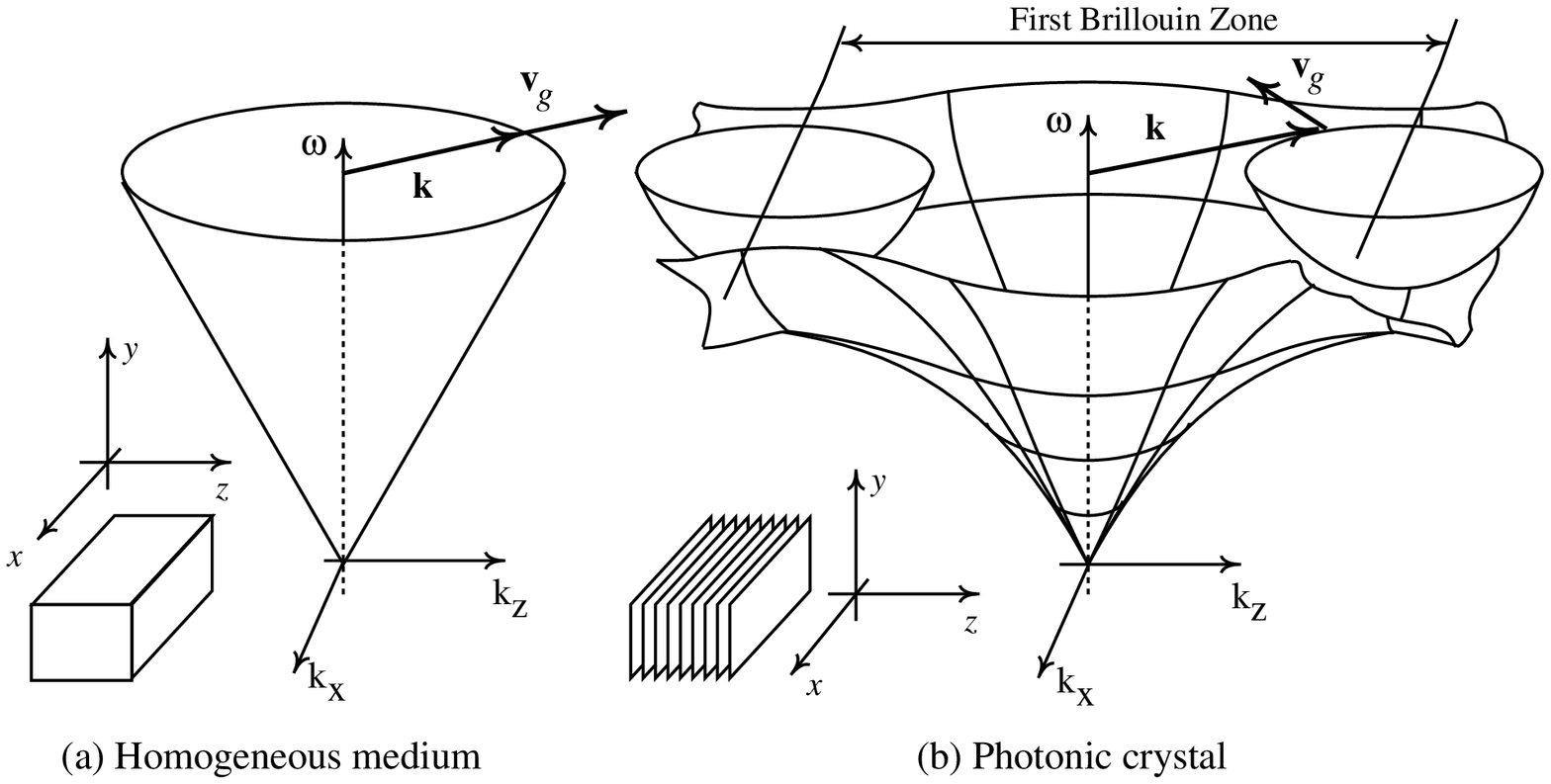} 
\begin{figure}
\caption{A 3D representation of the photonic band structures of (a) an isotropic homogeneous
nondispersive medium and (b) a 1D photonic crystal. Only 2D slices of the wave
vector space are depicted. Insets show the orientation of the media. A photonic
crystal band structure (b) is presented only for one basic polarization.\label{cones}}
\end{figure}
 
\noindent \epsfbox{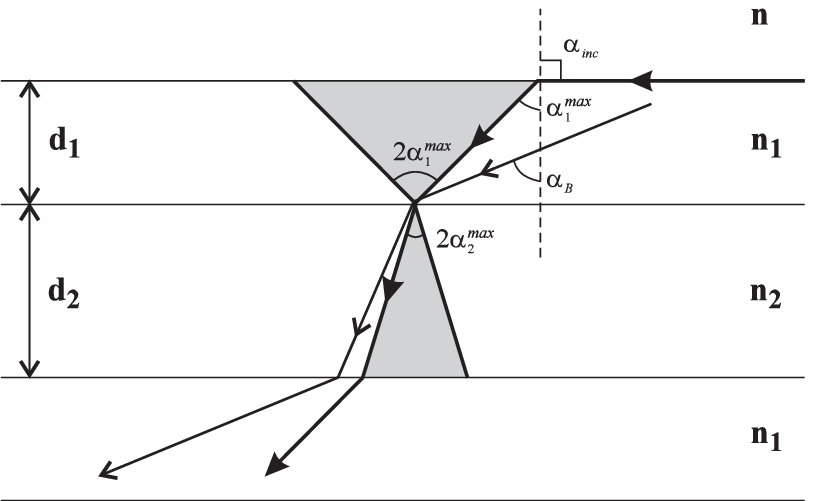} 
\begin{figure}
\caption{Schematic representation of a dielectric multilayer structure. The light rays
refracting and propagating through a stack are shown. The full domain of incident
angles \protect\( \alpha _{inc}\protect \) in the range from \protect\( -\pi /2\protect \)
to \protect\( \pi /2\protect \) is mapped onto the internal cone of half-angle
\protect\( \alpha ^{max}_{1}=\arcsin n/n_{1}\protect \) (light gray area).\label{Rays}}
\end{figure}

\noindent \epsfbox{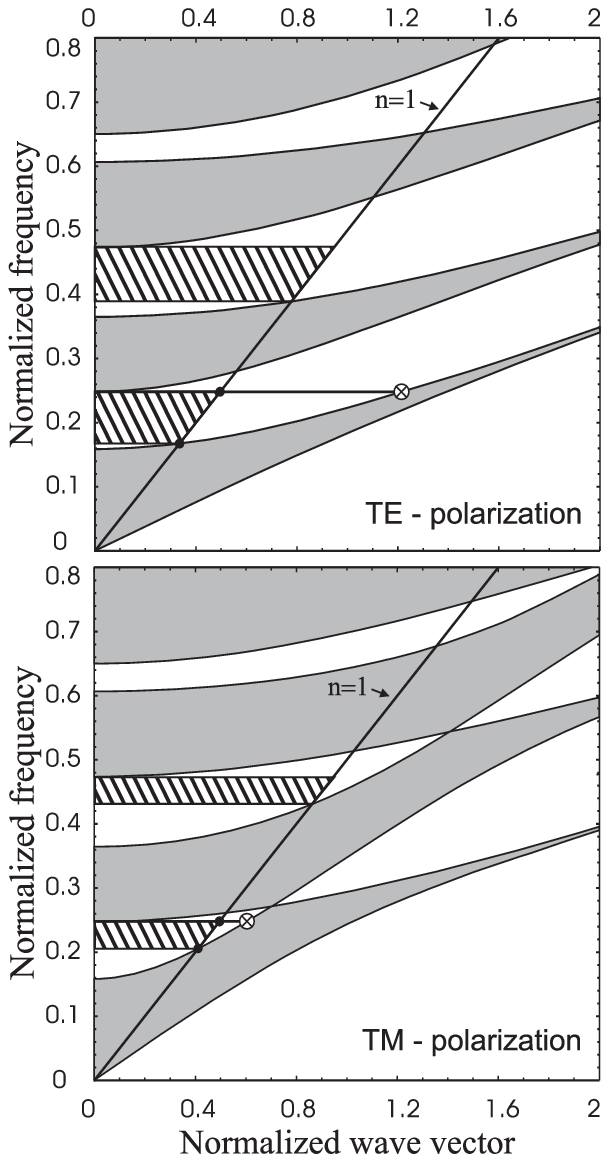} 
\begin{figure}
\caption{Projected band structure of a typical 1D photonic crystal for TE (top panel)
and TM (bottom panel) polarizations. The frequency and the tangential component
of the wave vector are defined to be normalized as \protect\( \omega \Lambda /2\pi c\protect \)
and \protect\( \left| {\textbf {k}}_{\perp }\right| \Lambda /\pi \protect \),
respectively. The gray areas correspond to the propagating states, whereas white
areas contain the evanescent states only. The shaded areas correspond to omnidirectional
reflection bands. The solid lines are the ambient-medium light-lines.\label{PBGk}}
\end{figure}

\noindent \epsfbox{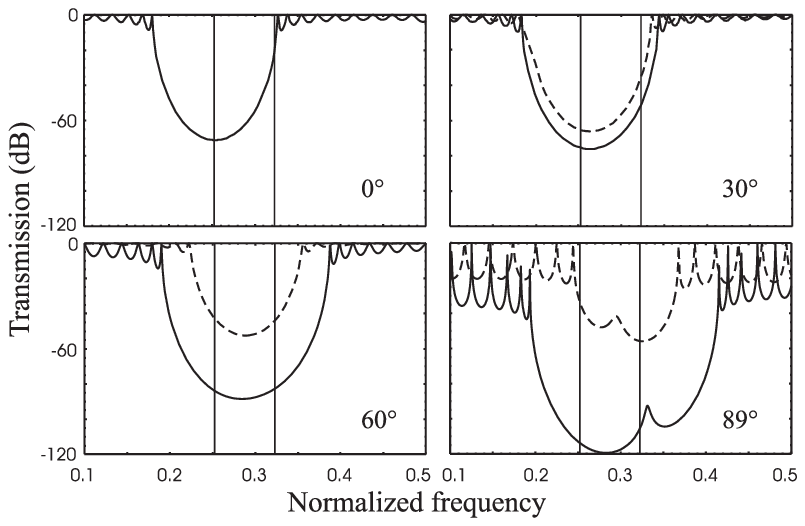} 
\begin{figure}
\caption{Transmission spectra from a Si/SiO\protect\( _{2}\protect \) structure with
10 periods at different incident angles, 0\protect\( ^{\circ }\protect \),
30\protect\( ^{\circ }\protect \), 60\protect\( ^{\circ }\protect \), 89\protect\( ^{\circ }\protect \).
The solid (dashed) curves are for TE (TM) polarization. The vertical lines on
each spectrum are the boundaries of a spectral region when omnidirectional reflection
occurs.\label{trans}}
\end{figure}

\noindent \epsfxsize=8.5cm \epsfbox{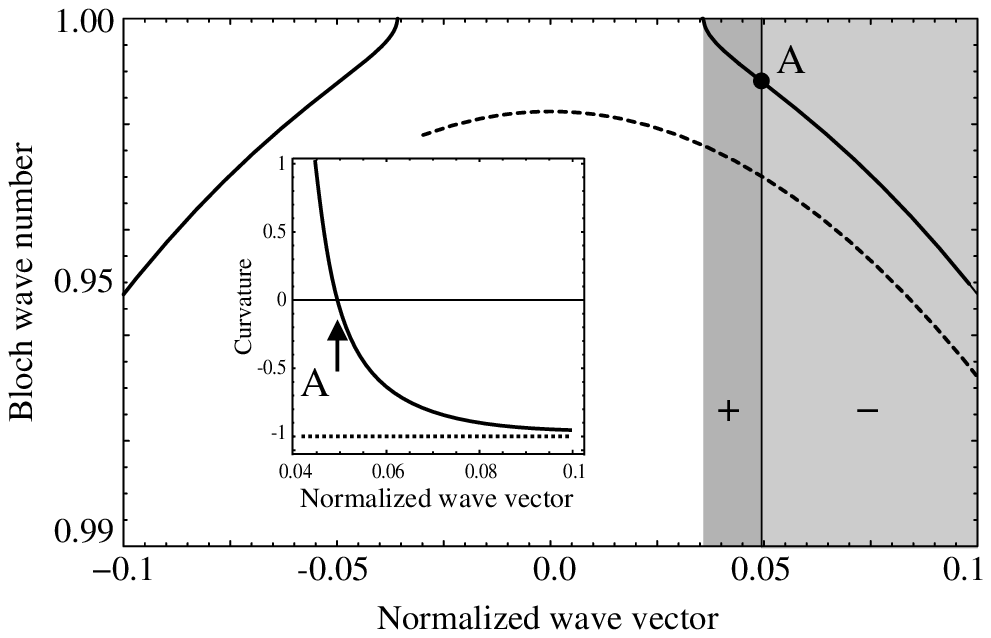} 
\begin{figure}
\caption{\protect\( K-k_{x}\protect \) intersection of dispersion surfaces of homogeneous
isotropic (dashed curve) and slightly modulated periodic (solid curve) media.
Normalized frequency is \protect\( \omega =0.146\protect \). Layers are of
equal thickness. Inset shows a local curvature of the dispersion contour of
photonic crystal. Arrow marks a wave vector for which the curvature is vanishing.
Concave and convex region of constant-frequency contours are marked in different
shades. Vertical and horizontal axes are not to scale.\label{ds_Caustics}}
\end{figure}

\noindent \epsfbox{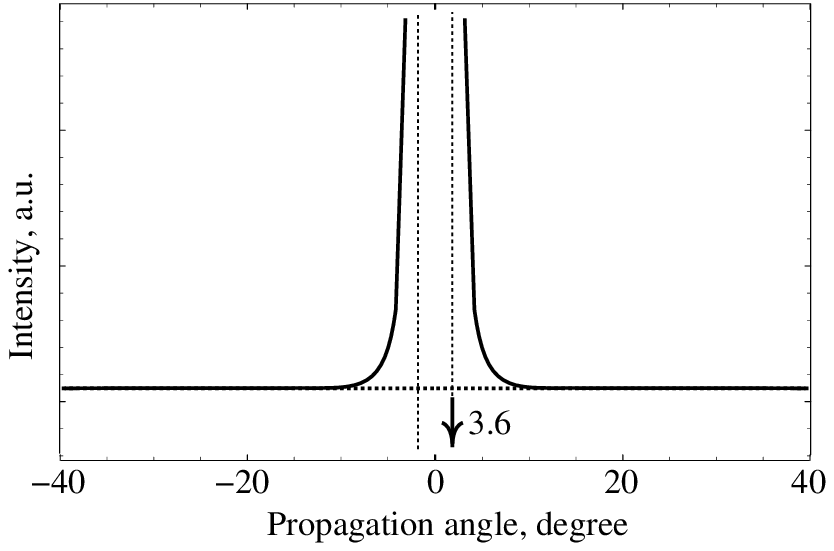} 
\begin{figure}
\caption{Angular intensity distributions. Solid curve, photonic crystal, shows an intense
peake in the direction corresponding to the inflection point. Dashed curve is
an isotropic intensity distribution corresponding to a homogeneous medium.\label{int_Caustics}}
\end{figure}

\end{document}